\documentclass[referee]{aa}
\begin{document}

   \thesaurus{06     
              (03.20.8;  
         	03.13.5;  
		11.11.1;) 
   \title{Interpreting the kinematics of the extended gas 
in distant radiogalaxies  from 8-10m telescope spectra.}
}

   \author{M. Villar-Mart\'\i n,
          \inst{1}
          A. Alonso-Herrero, \inst{1}
	 S. di Serego Alighieri \inst{2}
	  \and
	J. Vernet \inst{3}
          }

   \offprints{M. Villar-Mart\'\i n (mvm@star.herts.ac.uk)}

   \institute{Dept. of Natural Sciences, Univ. of Hertfordshire, College Lane, 
	Hatfield, Herts AL10 9AB, UK
	\and
	Osservatorio Astrofisico di Arcetri, Largo E. Fermi 5, I-50125, Firenze, Italy
	        \and
	European Southern Observatory,  Karl Schwarschild Str. 2, D-85748 Garching, Germany}

 \date{}

\titlerunning{Gas kinematics in distant radio galaxies}
\authorrunning{Villar-Martin et al.}
   \maketitle

   \begin{abstract}

	The nature of the extreme  kinematics in the extended gas of distant
radio galaxies ($z>$0.7) 
is still an open question. 
With the advent of the 8-10~m  telescope generation and the development
of NIR arrays we are in the
position for the first time to develop a more detailed study by using  lines other than
Ly$\alpha$ and [OII]$\lambda$3727 depending on redshift. 
In this paper we review the main sources of uncertainty 
 in the interpretation of the emission line kinematics: the presence of
several kinematic components,  Ly$\alpha$ absorption by neutral
gas/dust and the contribution of scattered light to some of the lines. As an example,
  several kinematic components can produce apparent, false rotation curves.
We propose methods to solve these uncertainties.

We propose to extend the  methods applied to low redshift radio galaxies
 to investigate the nature of the kinematics in distant radio galaxies: 
by means of the spectral decomposition of the strong optical emission lines 
(redshifted into the NIR) we can isolate
the different kinematic components  and  study  the {\it emission
line ratios} for the individual components.  If shocks are responsible for the
extreme kinematics, we should be able to isolate a kinematic  component 
(the shocked gas) with 
large FWHM ($\geq$900 km s$^{-1}$), low ionization level 
[OIII]$\lambda$5007/H$\beta \sim$2-4  and weak HeII$\lambda$4686/H$\beta\leq$0.07, 
together with a narrow
component ($\leq$few hundred km s$^{-1}$) with higher ionization level and strong
HeII emission (HeII/H$\beta\sim$0.5)

   \end{abstract}

%

\section{Introduction}
	
	High redshift radio galaxies ($z\geq$0.7, HzRG) show 
optical   regions of ionized
gas that extend across tens (sometimes hundreds)
 of kiloparsecs. These structures are
aligned with the radio axis (McCarthy et al. \cite{macc87} , Chambers et al. 
\cite{chamb87}) and show very irregular and clumpy 
morphologies.  The kinematics is extreme
 compared to the low redshift counterparts. 
Line widths (Ly$\alpha$ or [OII]$\lambda$3727) 
as broad as FWHM$\geq$1000 km s$^{-1}$
are often measured (McCarthy et al. \cite{macc96}, Villar-Mart\'\i n, Binette \& Fosbury \cite{villar99}, 
Baum \& McCarthy 
\cite{baum00}) 
versus $\sim$300-400  km s$^{-1}$ commonly observed at low
redshift ({\it e.g.} Tadhunter et al. \cite{tad89}). Velocity dispersions range from several
 hundreds up to
1600 km s$^{-1}$   and there are velocity
shifts between different lines 
of up to $\sim$1000 km s$^{-1}$  (R\"ottgering et al. \cite{rot97}).
Baum \& McCarthy (\cite{baum00}) showed that the transition from predominantly
quiescent systems to those with extreme motions occurs near $z\sim$0.6.

While at low redshift the kinematics of the extended gas in radio galaxies is
generally explained in terms of gravitational motions, the mechanism
responsible for the extreme motions at higher redshifts is not well
understood. The apparent connection between the size of the radio source and the emission
line kinematics 
observed at $z\geq$2   (van Ojik et al. \cite{ojik97}) suggests 
interactions between the radio and optical structures  that perturb the kinematics. 
This is supported by the
discovery in some HzRG of haloes of ionized gas extending {\it beyond} 
the radio structures that emit {\it narrow} Ly$\alpha$ ($\sim$ 250 km s$^{-1}$)
compared to the inner regions ($\sim$1200 km s$^{-1}$) (van Ojik et al. \cite{ojik96}). 

	Baum \& McCarthy (\cite{baum00}) studied a much larger sample covering a wider
range in redshift. They
 find {\it no} (or very weak) correlation between the emission line kinematics and 
the {\it ratio} of the radio to nebular
size. 
The authors support a gravitational origin for the kinematics of HzRG. Bipolar outflows 
(possible consequence of a circumnuclear starburst) have also
been proposed by some authors ({\it e.g.} Chambers \cite{cham98}, Taniguchi \& Shioya
\cite{tan00}).

	Understanding the nature of the kinematics in HzRG will help to answer
some open questions related to these galaxies:

\begin{itemize}

\item What is the nature of the alignment effect? 
Scattered light from
the hidden AGN is an important contributor ({\it e.g.} Cimatti
et al. \cite{cim97},\cite{cim98}, Fosbury et al. \cite{fos99}). 
However, the likely influence of radio jet/gas interactions is still not
understood. 

\item The ionization processes in the extended gas of  HzRG is clearly 
related to the nuclear activity, but what is the main mechanism:  jet driven shocks or
AGN illumination? 

\item  How does the radio source interact with its environment and how does it
influence the subsequent evolution of the system?

\item  Are very distant radio galaxies ($z\geq$2) in proto clusters?

\end{itemize}

To date, the kinematic studies of the gas in HzRG have been done in the optical.
The main limitation  has been  the need for a large
 collecting area to detect the emission lines with high signal/noise ratio (S/N) in
the extended gas. With 3-4m telescopes, only [OII]$\lambda$3727 ($z<$1.2) or Ly$\alpha$ ($z\geq$2) could
be used.
At ($z\geq$2) the uncertainties are important, since Ly$\alpha$ is highly sensitive
to dust/neutral gas absorption. The strong optical (rest frame)
emission lines ([OIII]$\lambda$5007,4959, [OII]$\lambda$3727)
are more reliable. However, they are redshifted into the NIR and the need for  a large collecting area
has constrained  the 
NIR spectroscopic studies of HzRG (and quasars) to  the spatially
integrated properties ({\it e.g.} Jackson \& Rawlings \cite{jack97}, 
McIntosh et al. \cite{mcin99}, Larkin et al. \cite{lar00})

\begin{figure*} 
\includegraphics{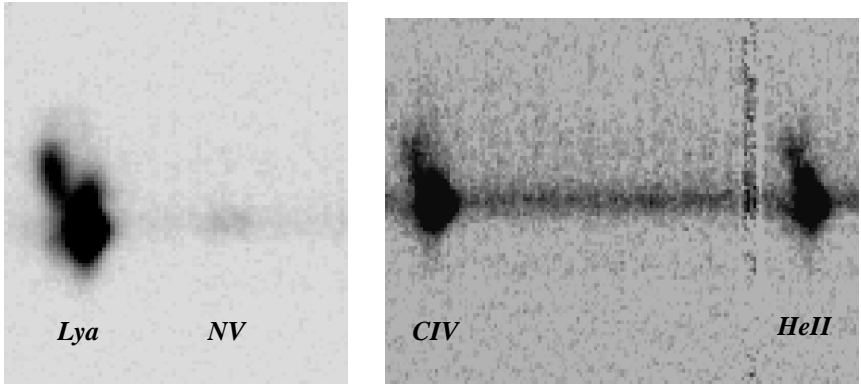}
\vspace{2.5in}
\caption{KeckII+LRISp spectra of the HzRG B3 0731+438 ($z=$2.42). The new generation of 8-10 m telescopes
allows for the first time the detection of emission lines other than Ly$\alpha$ 
in the extended gas of very distant radio galaxies ($z\geq$2) with high S/N.}
\label{Fig1}
\end{figure*}

This is the epoch of the 8-10 meter telescope generation.
We are in the position
for the first time to study the kinematics of the extended gas in HzRG:

\begin{itemize}

\item  using  UV rest frame lines  (redshifted into the optical) other than Ly$\alpha$  
 such as HeII,
CIV or CIII] (see Fig. \ref{Fig1}).

\item  using the optical rest frame lines such us [OIII]$\lambda$5007,4949,
H$\alpha$, etc (redshifted into the NIR).   The  development of   NIR arrays 
(and large collecting areas) makes this study possible for the first time.

\end{itemize}

In spite of the   opportunities opened by the new technological facilities,
the  study of the kinematic properties of HzRG is still complex.  The goal
of this paper is to
assess the main sources of uncertainty  and provide solutions to
solve them.

\section{The data}


%

	We present for illustrative purposes spectra of several 
high redshift radio galaxies ($z\sim$2.5) obtained with
the Low Resolution Imaging Spectrometer (LRISp, Oke et al. \cite{oke95}) at the 
Keck II telescope. Detailed description of the observing runs and data 
reduction will be presented in Vernet et al. (2001, in prep.).

\section{Uncertainties on interpreting the kinematic properties
of the extended gas in HzRG}

	The strongest lines in the optical (observed) spectra of HzRG
 are Ly$\alpha$,
CIV$\lambda$1550, HeII$\lambda$1640 and CIII]$\lambda$1909, while
[OII]$\lambda$3727, [OIII] $\lambda\lambda$5007,4959, H$\alpha$
dominate  
in the NIR.  There
are some physical effects 
that might make the interpretation of the line profiles and velocity
shifts uncertain.

-  Presence of multiple kinematic components 

- Broad scattered lines

- Resonant scattering of Ly$\alpha$ and (maybe)  CIV photons

\subsection{Multiple kinematic components}

We mentioned in \S1 that the interaction between the radio
jet and the ambient gas could be responsible for the extreme motions
in HzRG. Other authors favour a gravitational origin for the kinematics
of the gas.  One interpretation or another has important consequences:
if the velocity fields reflect the gravitational potential,
the derived  dynamical
mass  turns out to be correlated with redshift and/or radio power
(Baum \& McCarthy \cite{baum00}). As the authors pointed out, this
result is not valid if shocks are responsible for the kinematics.

Villar-Mart\'\i n  et al. (\cite{villar99b}) 
carried out a detailed spectroscopic study of the 
intermediate redshift radio galaxy PKS2250-51 ($z=$0.31) where a strong
interaction between the radio and optical structures occurs.
The authors resolved two main kinematic components, spatially extended
and detected in all optical (rest frame) 
emission lines. 
One of the components is 
narrow  ($\sim$150-200 km s$^{-1}$), the line ratios are consistent
with photoionization and it extends beyond the radio
structures. The other component is broad (FWHM as large as 900 km s$^{-1}$
at some spatial positions),
 the line ratios are consistent with shock ionization and it is emitted
 inside the radio structures. The properties of the
 narrow component  suggest that it is emitted by
ambient photoionized gas that has not been perturbed, while the properties
of the  broad component are consistent with shocked gas.
 
  A similar spectroscopic study
could prove or disprove
shocks as responsible for the  kinematics in HzRG. The studies done so far
 have been based  on 
the emission line kinematics along the radio axis and they 
  have not provided a definitive  answer.  We propose that the joint 
study of the line kinematics 
(with spectral decomposition of the line profiles) 
and the line ratios may give the answer. If shocks perturb
the kinematics in HzRG, then we should expect similar components 
and with similar flux ratios 
as those observed in PKS2250-41. With 2D spectrographs it will be possible to extend the
kinematic studies to regions far from the radio axis, where 
the interactions are expected to be non existent. The detection of
broad lines far from the radio axis would confirm that jet/cloud interactions
are not responsible for the extreme motions.

	The spectral resolution we use is  crucial to be able to
isolate different kinematic components in the extended gas.
A velocity resolution of
 $\sim$250 km s$^{-1}$ ($R=\lambda /\Delta\lambda \sim$1200)
 would be ideal since it will allow an accurate
decomposition of the emission line profiles.  The instrumental profile (IP) is in this
case well matched
with  the expected FWHM of the non
perturbed gas (so that we avoid unnecessary instrumental broadening), such 
as the diffuse haloes extending beyond the radio 
structures. The study of  such haloes
can be very useful, since they show the gas properties before  any
perturbation.  
With the new 1024$\times$1024 NIR arrays  it should be possible
to cover the interesting spectral range (HeII$\lambda$4686 to [OIII]$\lambda$5007) 
for a good number of objects.

\subsubsection{Apparent rotation curves}

	The presence of several kinematic components can lead us to
derive false rotation curves.

\begin{figure*} 
\includegraphics{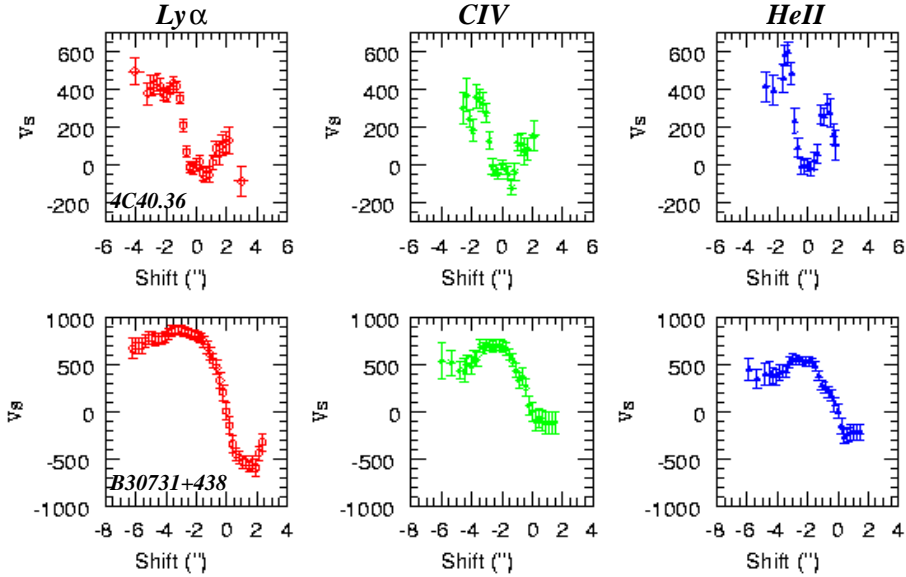}
\vspace{3.4in}
\caption{Velocity curves of Ly$\alpha$, CIV and HeII 
emission lines in
 4C40.36 ($z=$2.27) (top panels) and B3 0731+438 ($z=$2.43) (bottom panels). 
The spatial zero has been defined at the position of the
centroid of continuum emission in both cases. These velocity curves suggest rotation
of the ionized gas.}
\label{Fig2}
\end{figure*}

\begin{figure} 
\includegraphics{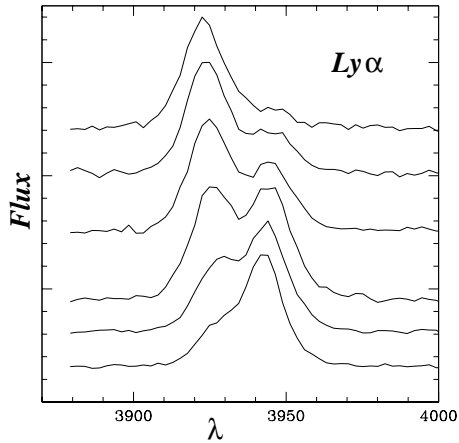}
\vspace{2.2in}
\caption{Spatial variation of the spectral profile of Ly$\alpha$ 
 in B3 0731+438. Each curve is the Ly$\alpha$
spectral profile (redshifted) at a given spatial position (pixel). 
Two kinematic components are present
whose relative contributions change from pixel to pixel.}
\label{Fig3}
\end{figure}

\begin{figure} 
\includegraphics{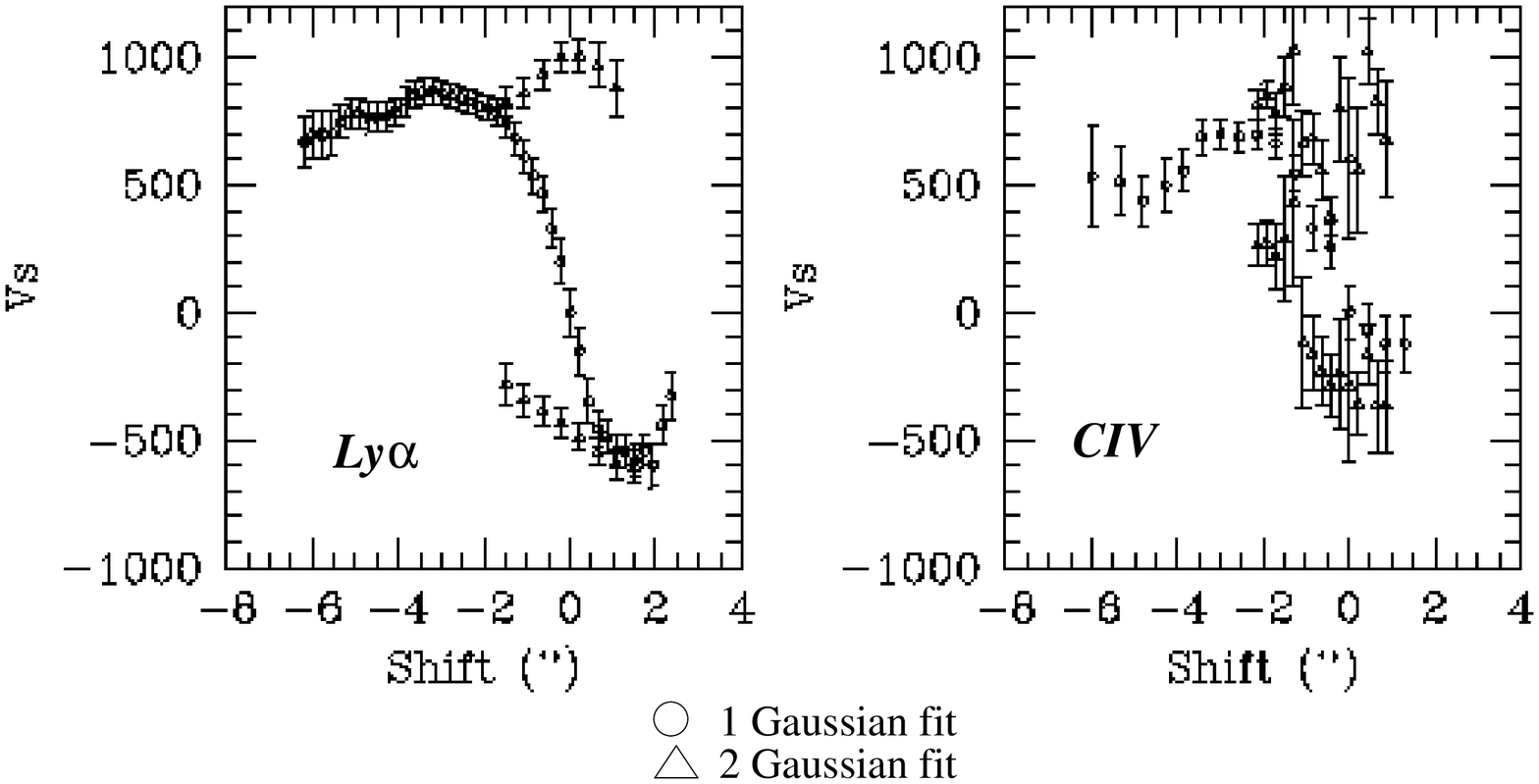}
\vspace{2.4in}
\caption{Comparison between the results obtained by
fitting 1 and 2 Gaussians to the spectral profile of the lines. 
The individual kinematic components are  shown  as open triangles 
and the result of fitting a single Gaussian is shown as open circles. 
 The gradient in the velocity curve
 is due to the change on the
 contribution of one  component relative to the other. The "apparent" 
rotation curve is an artifact consequence of the blend of both components
in some pixels.}
\label{Fig4}
\end{figure}

	Fig. \ref{Fig2} shows two examples of HzRG where the emission lines
show a resemblance with rotation curves (see also Fig. \ref{Fig1} and Fig. 2 in Villar-Mart\'\i n,
Binette \& Fosbury \cite{villar99}).  
This could be an exciting evidence
for  merger events ({\it e.g.} Hernquist \cite{her93}). However, 
the apparent rotation  is an artifact, consequence of the 
presence of at least two kinematic components. To illustrate this, we
have used the spectrum of the radio galaxy B3 0731+438 (Fig. \ref{Fig1}).
 We have extracted 1-D
spectra from  those spatial pixels where the abrupt jump in the apparent rotation curve
occurs.
Fig. \ref{Fig3} shows the Ly$\alpha$ spectral profile  at each pixel. The profile 
changes dramatically due to the presence of at least two kinematic components
 that contribute with different
relative intensity at different pixels. Fig. \ref{Fig1} shows clearly the two main
components. The fact that they are apparent also in HeII proves that it
is not an effect of resonant scattering of Ly$\alpha$ and CIV.
This makes the Ly$\alpha$ velocity centroid change in space, giving the
appearance of a rotation curve. If we isolate the  two components 
at every spatial position the rotation curve disappears (see Fig. \ref{Fig4}). 
 \footnote{Misleading velocity curves can also be a result of instrumental effects.
If  the seeing disc is noticeably smaller than the slit width 
and our object is clumpy (which is the case of HzRG) 
the clumps occupy different
spatial  positions in the direction perpendicular to the slit.
The centroid of the emission lines will be shifted in the spectral
direction. Another negative effect is that  the sky or arc lines may 
not represent the true instrumental profile and therefore we need to derive it
from the seeing value or, better, from  a point
source along the slit.}

\subsection{Broad scattered lines}

 Many HzRG are highly polarized in the optical  due to the scattering of nuclear emission (both continuum and emission lines
from the broad line region) by (probably) dust in the extended gas 
({\it e.g.} Cimatti
et al. \cite{cim97},\cite{cim98}, Fosbury et al. \cite{fos99}).  The radiation from the extended gas is
therefore  a mixture of direct and scattered light. Can the scattered broad lines affect our conclusions on the
kinematic studies if neglected?

\begin{figure} 
\includegraphics{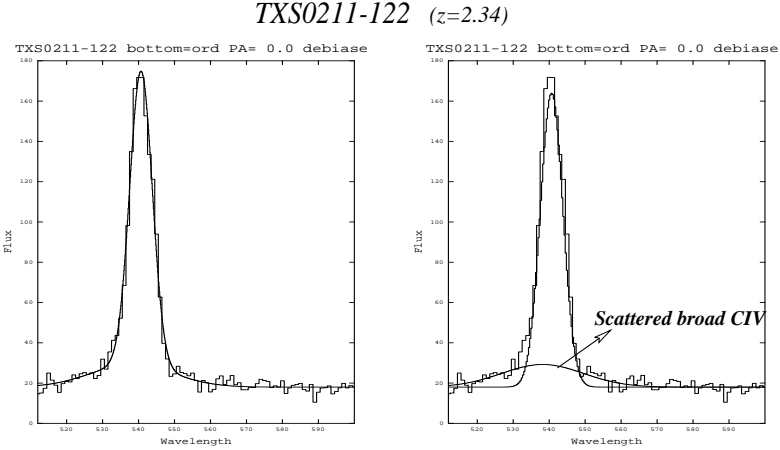}
\vspace{2.5in}
\caption{TXS0211-122 is one the most highly polarized HzRG. Although broad scattered CIV seems to be present,
it does not affect the narrow component, which dominates the emission.
Right panel: fit and data. Left panel: individual components of the fit
and data.}
\label{Fig5}
\end{figure}

As an example  we present 
 the spectrum of  TXS0211-122 ($z=$2.34), one of
 the most highly
polarized (P$\sim$20\% longward Ly$\alpha$) radio galaxies at high redshift. 
We have analysed the profile of  CIV,  which is efficiently
emitted in the broad line region and therefore is subject to scattering.
Fig. \ref{Fig5} shows the spectrum in the
region of CIV. There is an underlying broad component  with FWHM$\sim$4000
km s$^{-1}$. Similar FWHM are often observed in high
redshift quasars suggesting that it is 
 scattered light.

	The fit shows that  the CIV line profile is not seriously affected
by the underlying broad component 
due to the prominence of the "narrow" emission.
This is usually the case (also for
Ly$\alpha$) and  the studies that have found FWHM$\geq$1000  km s$^{-1}$
in the extended gas of HzRG are not affected by the effects of scattered light  (McCarthy et al. \cite{macc96}, Villar-Mart\'\i n, Binette 
\& Fosbury  \cite{villar99}).
However, we cannot neglect this contribution when studying the line profiles in more detail
 (looking for different kinematic components, for instance); we
 would  interpret 
the broad scattered components 
as due to extreme motions in the extended gas.

The best way to avoid any possible uncertainty on this issue is 
 to compare with lines that are not emitted efficiently in the BLR
(and, therefore, they are not scattered), such as HeII$\lambda$1640 (Foltz et al.
\cite{foltz88},
 Heckman et al.  \cite{heck91}) or 
 optical forbidden lines such as  [OII] and [OIII]$\lambda$5007,4959 (the [OIII] 
lines might have a minor contribution from scattered light, di Serego Alighieri et al.
 \cite{dise97}).

\subsection{Resonant scattering of Ly$\alpha$ and CIV}
	
	Ly$\alpha$ and CIV$\lambda$1550 are resonant lines and intervening
HI and CIV respectively can absorb the emission. Ly$\alpha$ is particularly sensitive
to this effect. The discovery of absorption troughs in the profile
of
the line across the whole extension of the Ly$\alpha$ emitting gas
 in many HzRGs  has proved that
Ly$\alpha$ absorption  is a common phenomenon (van Ojik et al. \cite{ojik97}) . 
  CIV absorption troughs have aslo been observed  (e.g. Binette et al.  \cite{bin2000}). 
Van Ojik et al. 
found a correlation between the radio size and the occurrence of absorption,
suggesting that in radio galaxies with large radio sources the effect
is less worrying.

	Absorption of Ly$\alpha$ photons
 can modify dramatically the profile of the line
(van Ojik et al. \cite{ojik97}). R\"ottgering et al. (\cite{rot97})
concluded that the effects of associated H I absorption 
 may be responsible for the shift of the Ly$\alpha$ line with respect
to the high ionization lines in some objects. 
By using non resonant lines (HeII$\lambda$1640, forbidden lines)
we will avoid
this problem. The CIV doublet, although resonant, is less sensitive and more reliable
than Ly$\alpha$. The use of high spectral resolution ($\sim$3 \AA) can
also help, since we will be able to resolve the absorption troughs 
(van Ojik et al. \cite{ojik97}).

\subsection{The line doublets}

	The comparison between the FWHM of the emission lines provides 
information about the processes responsible for the line emission and
the kinematics. The detection of different line velocity  widths for 
different lines will lead us to the conclusion that different mechanisms
are at work  and/or there are regions of different physical conditions. As an example,
some radio galaxies with radio/optical interactions show that low ionization
lines are broader than high ionization lines (Clark et al. \cite{clark98}). This
has been interpreted as a consequence of shocks: the shocked gas has lower
ionization level and more perturbed kinematics than the non perturbed gas. 

In this sense, care must be taken 
with the doublets (CIV$\lambda$1550, CIII]$\lambda$1909 and [OII]$\lambda$3727). At high redshifts
the separation  between the two components
is large (7 \AA ~ at $z=$2.5 and 10 \AA ~ at $z=$4) and
the doublet as a whole can show a profile apparently broader 
than  single lines. This will be the case when the lines are similar in (observed) width or narrower than the doublet separation.
If the  lines are much broader  the two components  will be severely  blended, whatever
the resolution we use and the broadening  will not be important. This should often  be the case
since gas motions can 
produce lines of intrinsic FWHM $\geq$20 \AA ~ at $z=2.5$..

\section{Optical vs. near infrared}

\begin{figure}
\includegraphics{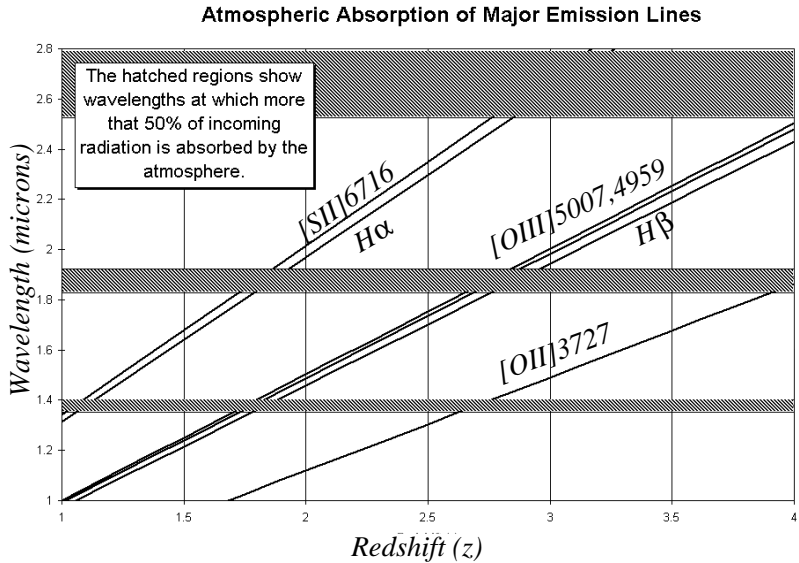}
\vspace{3in}
\caption{The inclined solid lines indicate the variation with redshift of the observed 
wavelength
for several important emission lines. The hatched regions show the spectral coverage
of the atmospheric bands where the absorption is more than 50\% of the incoming
radiation. The redshifts for which one of the emission lines lies in these dark regions
can be easily deduced from the plot.
}
\label{Fig6}
\end{figure} 
	The NIR spectral range offers some  advantages compared to the optical.
The lines (optical rest frame)  are less affected by dust 
extinction.  Moreover, we can use the same
 lines that have provided most of the information about low redshift objects.
The comparison is therefore easier and more reliable than comparing the UV rest frame
at high redshift with the optical rest frame at low redshift. 
We can also use the 
  diagnostic tools developed for low redshift objects to investigate  
the physical conditions  and ionization state of the extended gas. 
Such diagnostic techniques are reasonably well understood, contrary 
to what happens
in the UV rest frame.

	The main problems we may find when observing in the NIR are
due to the effects of the atmosphere and the thermal emission from
the telescope.  Due to atmospheric absorption some redshifts are forbidden to study certain
emission lines (see Fig. \ref{Fig6}).

 On the other hand, HzRG
are very faint sources and the  OH bands (dominant source of sky emission in the
NIR) are bright and variable compared to
them so that in same cases it will be impossible to obtain enough S/N on the line
of interest. Since the FWHM of the object lines is large (e.g. [OIII]$\lambda$5007$\sim$40-50 
\AA ~ at $z=$2.5),  the  contamination will often be important 
independently of the spectral resolution we use, unless we work with  emission lines
that lie far from strong sky lines.  This puts constrains on the object
redshifts.
 At longer wavelengths ($\geq$2.2 $\mu$m) the limitations are due to  thermal 
background.

\section{Summary and conclusions}

	We have discussed the possible uncertainties and proposed solutions for the
interpretation of the kinematics of the extended gas in HzRG using  the
optical (UV rest frame) and NIR (optical rest frame)  emission lines.

\begin{itemize}

\item  The spectral decomposition of the emission lines 
 will  show multiple  components.
By studying the kinematic, flux and spatial properties of the individual 
components we can disentangle 
the mechanisms responsible for the extreme motions.

\item The presence of several kinematic components (or absorption in the
case of Ly$\alpha$) may produce apparent rotation curves. 

\item   Ly$\alpha$  aborption by neutral gas/dust  can
change dramatically the emission line profiles. CIV$\lambda$1550 is also resonant, but
less sensitive. The detection of  non 
resonant lines   with good S/N is important to understand the 
effects of resonant scattering.

\item  When scattered broad lines are present, they do not affect the emission line
 profiles of
the brightest lines. However, they must be taken  into account when studying
different kinematic components, otherwise  high velocity motions in the extended gas
could be inferred. HeII$\lambda$1640 and the forbidden optical lines can help
to determine the contribution of scattered light.  

\item The doublets  (CIV, CIII], [OII])  are 
noticeably broader than simple lines (like HeII) when the gas motions have
dispersion velocities similar to the separation between the doublet components.

\item  NIR observations are better suited for kinematic 
studies of HzRG since    they map optical rest frame lines that 
are less sensitive to dust and
the effects mentioned above. An additional advantage is that
this spectral range allows the comparison 
with studies 
 at low redshift  so that we can use  powerful diagnostics tools developed for
nearby objects.  However, the atmosphere sets 
limitations to this type of observations.

\end{itemize}

\begin{acknowledgements}
     The authors thank the members of the `z2p5' collaboration A. Cimatti, M. Cohen, 
B. Fosbury and 
B. Goodrich   for the data presented in this
paper for illustrative purposes.  We thank
  Phil Lucas for useful discussions and Callum McCormick for providing Figure 6.
\end{acknowledgements}


\begin{thebibliography}{}

 \bibitem[2000]{baum00} Baum S., McCarthy P., 2000, AJ 119, 2634

 \bibitem[2000]{bin2000}Binette L., Kurk J., Villar-Mart\'\i n M.,  R\"ottgering H.,
2000, A\&A 356, 23

\bibitem[1987]{chamb87} Chambers K.C., Miley G.K., van Breugel W., 1987, Nature 329, 604  

\bibitem[1998]{cham98} Chambers K.C., 1998, in: American Astronomical Society
 Meeting,
193, 110.01

\bibitem[1997]{cim97} Cimatti A., Dey A., van Breugel W., Hurt T., Antonucci R., 1997, ApJ 476, 677

\bibitem[1998]{cim98} Cimatti A., di Serego Alighieri S., Vernet J., Cohen M., Fosbury R., 
1998, ApJ 499, 21

    \bibitem[1994]{cor94} Corbin M., Francis P., 1994, AJ 108, 2016 


\bibitem[1998]{clark98} Clark N.E., Axon D., Tadhunter C.N., Robinson A.,  O'Brien P., 1998, ApJ 494, 546

 \bibitem[1997]{dise97} di Serego Alighieri S., Cimatti A., Fosbury R., Hes R., 1997,
A\&A 328, 510

  \bibitem[1998]{foltz88} Foltz C., Chaffee F., Weymann T., Anderson S., 1988, in {\it QSO Absorption
Lines: Probing the Universe}, ed. J.C. Blades, D. Turnshek \& C. Norman (Cambridge: Cambridge University
Press), p.\ 53

 \bibitem[1999]{fos99} Fosbury R., Vernet J., Villar-Mart\'\i n M.,  Cohen M., Cimatti A., 
di Serego Alighieri S., McCarthy P., 1999, in {\it ESO Conference on Chemical Evolution from Zero
to High Redshift}, Garching, Germany, October 14-16 1998. ESO Astrophysics
Symposia, Eds. J. Walsh and M. Rosa, Springer



  \bibitem[1991]{heck91}  Heckman T., Lehnert M., Miley G., van Breugel W., 1991, ApJ 381, 373

\bibitem[1993]{her93} Hernquist L., 1993, ApJ 409, 548

 \bibitem[1997]{jack97} Jackson N.,  Rawlings S., 1997, MNRAS 286, 241

 \bibitem[2000]{lar00}  Larkin J. et al., 2000, ApJL 533, 61

 \bibitem[1987]{macc87}  McCarthy P.J., Spinrad H., Djorgovsky S.,
Strauss M.A., van Breugel W., Liebert J., 1987, ApJ 319, L39

\bibitem[1996]{macc96}  McCarthy P.J., Baum S., Spinrad H., 1996, ApJS 106, 281

 \bibitem[1999]{mcin99} McIntosh D., Rieke M., Rix H., Foltz C., Weymann J., 1999, ApJ 514, 40

 \bibitem[1995]{oke95} Oke  et al, 1995, PASP 107, 375

 \bibitem[1997]{rot97} R\"ottgering H.,  van Ojik R., Miley G., Chambers K., van Breugel W., de Koff
S., 1997, A\&A 326, 505  

 \bibitem[1989]{tad89} Tadhunter C., Fosbury R., Quinn P., 1989, MNRAS 240, 225

 \bibitem[2000]{tan00} Taniguchi Y., Shioya Y., 2000, ApJL 532, 13

 \bibitem[1996]{ojik96} van Ojik R., R\"ottgering H., Carilli C.,
 Miley G., Bremer M., Macchetto F., 1996, 
A\&A 313, 25

 \bibitem[1997]{ojik97} van Ojik R., R\"ottgering H., Miley G., Hunstead R., 1997,
A\&A 318, 358


 \bibitem[1996]{villar96} Villar-Mart\'\i n M., Binette L., Fosbury R., 1996, A\&A 312, 751

  \bibitem[1999]{villar99} Villar-Mart\'\i n M., Binette L., Fosbury R., 1999, 
A\&A 346, 7

 
  \bibitem[1999]{villar99b} Villar-Mart\'\i n M., Tadhunter C., Morganti R., 
Axon D., Koekemoer A., 1999, MNRAS 307, 24




\end{thebibliography}
\end{document}